\documentclass[onecolumn,showpacs]{revtex4}

\usepackage{amsmath}
\usepackage{graphicx}
\usepackage{dcolumn}
\usepackage{bm}

\input epsf

\begin{document}

\title{\Large Thermodynamic Prescription of Cosmological Constant in Randall Sundrum-II
Brane}

\author{\bf Tanwi~Bandyopadhyay\footnote{tanwi.bandyopadhyay@aiim.ac.in}}

\affiliation{Adani Institute of Infrastructure
Engineering,~Ahmedabad-382421, India.}

\date{\today}

\begin{abstract}
In this work, we apply quantum corrected entropy function derived
from the Generalized Uncertainty Principle (GUP) to the
Holographic Equipartition Law to study the cosmological scenario
in Randall-Sundrum (RS) II brane. An extra driving term has come
up in the effective Friedmann equation for a homogeneous,
isotropic and spatially flat universe. Further, thermodynamic
prescription of the universe constraints this term
eventually with order equivalent to that of the cosmological constant.\\


\end{abstract}


\maketitle

\section{\normalsize\bf{Introduction}}

In order to give an explanation of higher dimensional theory,
Randall and Sundrum (\cite{RS1},~\cite{RS2}) proposed an idea of
bulk-brane model, where the four dimensional world in which we
live is called the 3-brane (a domain wall) that is embedded in a
higher dimensional spacetime (bulk). According to the theory, the
brane confines all the matter field, only gravity propagates in
the bulk. Moreover the extra fifth dimension need not be finite,
it can extend to infinity in either side of the brane. The concept
of brane world scenarios shows a possibility to resolve the
problem of unification of all forces and particles in nature. The
main equations governing the cosmological evolutions of the brane
differ from the corresponding Friedmann equations in standard
cosmology (\cite{BDEL1} -\cite{Ponce2}). The difference lies in
the fact that the energy density of the brane appears to be in a
quadratic form whereas in standard cosmology, the energy density
appears linearly in the field equations. This model is also
consistent with the string theory and may resolve the so called
hierarchy problem or the source of dark energy and dark matter
(\cite{Lucas1} ,~\cite{Lukas2}). The later theory is one of the
overwhelming theories of the current era. The concept of dark
matter had been first proposed (\cite{Zwiecky1},~\cite{Zwiecky2})
in the context of studying galaxy clusters. The dark energy, on
the other hand, is a completely new component which produces
sufficient negative pressure. This drives the cosmic acceleration
which has also been substantiated by the observational evidences
over the years. The observational data clearly states that the
current universe is flat having approximate cosmic content of 21\%
dark matter, 72\% dark energy and rest in the form of visible
matter and radiation. All these imply that the standard
cosmological models are needed to be modified with the models of
dark matter and dark energy. Unfortunately, very less is known
about dark energy. Hence there exist many prospective candidates
for this cosmic component. Among them, cosmological constant
$\Lambda$ is the most popular having an equation of state
$p_\Lambda=-\rho_{\Lambda}$. This model is known as $\Lambda$CDM
model (cold dark matter) (\cite{Perl}-\cite{Ade}). This theory has
a major drawback in terms of order of measurement. The observed
value of $\Lambda$ is many order of magnitude smaller than its
theoretical value predicted in quantum field theory. This is
termed as the cosmological constant problem and to resolve this,
one of the many proposed cosmological models is varying
cosmological constant ($\Lambda(t)$CDM) model (\cite{Shapiro}-\cite{Lima}).\\

On the other hand, one of the key features of quantum theory of
gravity is called the holographic principle. This states that in a
bounded system, the number of degrees of freedom is associated to
entropy and scales with the area enclosed
(\cite{'tHooft}-\cite{Bousso}). Under this principle, gravity is
shown to be an entropic force derived from the changes in the
Bekenstein-Hawking entropy (\cite{Paddy1}-\cite{Verlinde}).
Further, many studies focussed on derivation and investigation of
the Friedmann and acceleration equations in the background of
entropic cosmology (\cite{Sheykhi1}-\cite{Mitra}). Various forms
of entropy have been applied in these studies
(\cite{Easson1}-\cite{Tsallis}). In some of them, an extra driving
term is derived from entropic forces on the horizon of the
universe in order to explain its accelerated expansion. Intrigued
by the holographic principle, very recently Padmanabhan
(\cite{Paddy3}) proposed a different approach saying that the
cosmic space is emergent as the cosmic time progresses. It has
been termed as the holographic equipartition law. According to
this, the rate of expansion of the universe is related to the
difference between the surface degrees of freedom on the
holographic horizon and the bulk degrees of freedom inside.
Keeping this in the background, the cosmological equations were
derived and examined both in classical and modified theories of
gravity (\cite{Cai2}-\cite{Moradpour}). For most of these studies,
Bekenstein-Hawking entropy played the major role.\\

Very recently, a similar study has been carried out in
\cite{Komatsu3}, where a modified R\'{e}nyi entropy was chosen
instead of Bekenstein-Hawking entropy and a constant like term was
obtained in the field equations. Imposing an analytical
constraint, this term showed behavior similar to the varying
cosmological constant. Further, the power-law corrected entropy
was also tested in the same mechanism and similar results were
found in \cite{Komatsu4}. This surely necessitates more
investigation to the alternative studies of dark energy and
cosmological constant in modified gravity theories. We have
followed this novel approach to study the underlying cosmological
scenario in the RS-II brane model considering the quantum
corrected form of the entropy function derived from the
Generalized Uncertainty Principle (GUP) \cite{Vagenas}. A similar
study has been carried out in \cite{Tu} in Einstein's gravity but
our entropy function in unique in its $\sqrt{Area}$ form. The
necessity and motivation for choosing this entropy function was
discussed later in details. The GUP corrected entropy was applied
to the holographic equipartition law in a four dimensional
universe embedded in a conformally flat five dimensional
space-time. Consequently, an analogous extra driving term is
derived in the modified Friedmann equations. Further
thermodynamical investigations showed that this extra term is of
order identical to the order of cosmological constant.\\

The paper is organized in the following way: In section II, we
briefly review the $\Lambda(t)$CDM model and the modified field
equations in the context of brane world gravity. In section III,
the expansion of the cosmic space is treated as an emergent
process and the modified Friedmann equations are retrieved from
the Holographic Equipartition Law in the absence of any dark
energy component. Section IV presents a brief review of GUP
corrected entropy. In this section, subsection A discusses the
results of application of GUP corrected entropy into the
holographic equipartition law. In subsection B, the validity of
the Generalized Second Law of Thermodynamics (GSLT) is assumed and
the behavior of the extra driving term is analyzed. Finally,
a brief discussion on our study is made in section V.\\

\section{Main Equations:~$\Lambda(t)$\bf{CDM Model in Brane World}}

A homogeneous, isotropic, spatially flat
Friedmann-Robertson-Walker (FRW) universe in the natural unit
system ($G=c=\hbar=k_B=1$) is given by

\begin{equation}
ds^2=dt^2 -a^2(t)\left[dr^2 +r^2(d\theta^2 +sin^2\theta
d\phi^2)\right]
\end{equation}

which is considered to be embedded in a conformally flat five
dimensional space-time. The form of the energy momentum tensor for
a combination of dark matter and dark energy is

\begin{equation}\label{momentum}
{{T_{\mu}}^{\nu}}=(\rho_m +p_m+\rho_{\Lambda}+p_{\Lambda})u_\mu
u^\nu -(p_m+p_{\Lambda}){{\delta_{\mu}}^{\nu}}
\end{equation}

Generally a barotropic equation of state $p_m=\omega_m \rho_m$ is
chosen for the matter part on the brane having energy density
$\rho_m$ and pressure $p_m$ and a variable cosmological constant
is chosen as the component of dark energy having energy density
$\rho_{\Lambda}$ and pressure $p_{\Lambda}(=-\rho_{\Lambda})$. The
four velocity $u_\mu$ in comoving coordinate system takes the form
$u_\mu={\delta_{\mu}}^t$. Thus the effective Einstein equations on
the brane are \cite{Tanwi}

\begin{equation}\label{field eqn}
\frac{{\dot{a}}^{2}}{a^{2}}= H^2= \frac{8\pi}{3}
\left[\rho_{T}\left(1+\frac{\rho_{T}}{2\lambda}\right)\right]
\end{equation}

and

\begin{equation}\label{accln eqn}
\frac{\ddot{a}}{a}=\dot{H}+H^2=-\frac{4\pi}{3}
\left[\rho_{T}\left(1+\frac{2\rho_{T}}{\lambda}\right)+3
p_{\Lambda}\left(1+\frac{\rho_{T}}{\lambda}\right)\right]
\end{equation}

where $\rho_{T}=\rho_m+\rho_{\Lambda}$ is the total energy
density, $p_{T}=p_m+p_{\Lambda}$ is the total pressure, $\lambda$
is the positive brane tension, the Hubble parameter is given by
$H(t)=\frac{\dot{a}}{a}$ and a(t) is
the scale factor in flat FRW brane model.\\

Equations \eqref{field eqn} and \eqref{accln eqn} can be
explicitly written as

\begin{equation}\label{mod fld eqn}
\frac{{\dot{a}}^2}{a^2}=\frac{8\pi}{3}{{\rho}_m}_{\text{eff}}+\frac{1}{3}\Lambda(t)_{\text{eff}}
\end{equation}

and

\begin{equation}\label{mod accln eqn}
\frac{\ddot{a}}{a}=-\frac{4\pi}{3}\left[(\rho_m+3p_m)+\frac{1}{\lambda}(2\rho_m+3p_m+2\rho_m
\rho_\Lambda+3\rho_m p_m)\right]
+\frac{1}{3}\Lambda(t)_{\text{eff}}
\end{equation}

where
${{\rho}_m}_{\text{eff}}={\rho}_m\left(1+\frac{{\rho}_m}{2\lambda}\right)$
and
$\Lambda(t)_{\text{eff}}=8\pi\left[{\rho}_{\Lambda}\left(1+\frac{{\rho}_{\Lambda}}{2\lambda}
+\frac{{\rho}_m}{\lambda}\right)\right]$.\\

For the present brane model with matter field given by equation
\eqref{momentum}, the explicit form of the energy momentum
conservation relation (${{T_{\mu}}^{\nu}}_{;\nu}=0$) is

\begin{equation}\label{mod cont eqn}
\dot{\rho_m} +
3\frac{\dot{a}}{a}(\rho_m+p_m)=-\dot{\rho_\Lambda}\simeq
-\frac{\dot{\Lambda(t)_{\text{eff}}}}{8\pi}
\end{equation}

Instead of a variable $\rho_{_\Lambda}$, if we choose a constant
$\rho_{\Lambda}$, then the field equations together with the
continuity equation will be identical to the corresponding
equations in the standard $\Lambda$CDM model.\\

\section{\normalsize\bf{Field equations derived from the Holographic Equipartition Law}}

For a pure de Sitter universe with Hubble parameter $H$, the
holographic principle can be described by the relation
\cite{Paddy3}

\begin{equation}
N_{\text{sur}}=N_{\text{bulk}}
\end{equation}

where $N_{\text{sur}}$ denotes the number of the degrees of
freedom on the holographic screen with Hubble radius $r_H=1/H$

\begin{equation}\label{N Sur}
N_{\text{sur}}=\frac{4\pi}{H^2}=4S_{H}
\end{equation}

Here $S_H$ is the entropy on the Hubble horizon. The number of
degrees of freedom in bulk is said to obey the equipartition law
of energy

\begin{equation}
N_{\text{bulk}}=\frac{2|E|}{T}
\end{equation}

In the context of brane world models, the induced active
gravitational mass on the brane $|M|=|E|$ has the form
\cite{Ling2}

\begin{eqnarray*}\label{Komar mass}
|M|=\frac{4\pi}{3H^3}\left|\left(\rho_T+3p_T+\frac{3\rho_T
p_T}{\lambda}+\frac{2{\rho_T}^2}{\lambda}\right)\right|
\end{eqnarray*}
\begin{equation}
~~~~~~~~~~~~~~~~~~~~~~~~~~~~~~~~~=-\epsilon
\frac{4\pi}{3H^3}\left\{\left[(\rho_m+3p_m)+\frac{1}{\lambda}(2\rho_m+3p_m+2\rho_m
\rho_\Lambda+3\rho_m p_m)\right]
+\frac{1}{4\pi}\Lambda(t)_{\text{eff}}\right\}
\end{equation}

for the choice of the matter field \eqref{momentum}. The parameter
$\epsilon$ is defined later. Using the above expression of $|M|$
and the horizon temperature $T=H/2\pi$, we get the expression of
$N_{\text{bulk}}$ as

\begin{equation}\label{N Bulk}
N_{\text{bulk}}=-\epsilon
\frac{16{\pi}^2}{3H^4}\left\{\left[(\rho_m+3p_m)+\frac{1}{\lambda}(2\rho_m+3p_m+2\rho_m
\rho_\Lambda+3\rho_m
p_m)\right]+\frac{1}{4\pi}\Lambda(t)_{\text{eff}}\right\}
\end{equation}

Since the real world is not purely but asymptotically de Sitter,
therefore one may propose that the expansion rate of the cosmic
volume is related to the difference of these two degrees of
freedom. The analytical form of this is described as\cite{Paddy3}

\begin{equation}\label{change V}
\frac{dV}{dt}={l_p}^2(N_{\text{sur}}-\epsilon N_{\text{bulk}})
\end{equation}

Equation \eqref{change V} is known as the holographic
equipartition law. Here $V=\frac{4\pi}{3H^3}$ is the cosmic volume
and the parameter $\epsilon$ is defined by
(\cite{Paddy3},\cite{paddy5})

\begin{equation}\label{defn}
\epsilon \equiv \left\{
\begin{array}{ll}
+1,~~~~~~\text{when}~~\left[(\rho_m+3p_m)+\frac{1}{\lambda}(3p_m+2\rho_m+3\rho_m p_m)\right]<0\\\\
-1,~~~~~~\text{when}~~
\left[(\rho_m+3p_m)+\frac{1}{\lambda}(3p_m+2\rho_m+3\rho_m
p_m)\right]>0
\end{array}
\right.
\end{equation}

Here, we have considered that there is no dark energy component in
the 3-brane, i.e, $\Lambda(t)_{\text{eff}}\sim \rho_\Lambda=0$. In
this case
$\left[(\rho_m+3p_m)+\frac{1}{\lambda}(3p_m+2\rho_m+3\rho_m
p_m)\right]<0$ for the acceleration of the universe. Hence from
equation \eqref{Komar mass} and \eqref{defn}, the definition of
the parameter $\epsilon$ is well justified.\\

One can write from equations \eqref{N Sur}, \eqref{N Bulk} and
\eqref{change V}

\begin{equation}
-4\pi
\frac{\dot{H}}{H^4}=\left\{4S_H+\frac{16{\pi}^2}{3H^4}\left[(\rho_m+3p_m)+\frac{1}{\lambda}(3p_m+2\rho_m+3\rho_m
p_m)\right]\right\}
\end{equation}

or equivalently

\begin{equation}
\dot{H}=-\frac{4\pi}{3}\left[(\rho_m+3p_m)+\frac{1}{\lambda}(3p_m+2\rho_m+3\rho_m
p_m)\right]-\frac{H^4 S_H}{\pi}
\end{equation}

The acceleration equation is therefore read as

\begin{equation}\label{new accln eqn}
\frac{\ddot{a}}{a}=-\frac{4\pi}{3}\left[(\rho_m+3p_m)+\frac{1}{\lambda}(3p_m+2\rho_m+3\rho_m
p_m)\right]+H^2\left(1-\frac{H^2 S_H}{\pi}\right)
\end{equation}

Thus we have derived the acceleration equation from the
holographic equipartition law and an extra driving term appears in
the right side of the equation. This term vanishes when one
chooses the Bekenstein-Hawking entropy for $S_H$. The acceleration
equation will then be

\begin{equation}
\frac{\ddot{a}}{a}=-\frac{4\pi}{3}\left[(\rho_m+3p_m)+\frac{1}{\lambda}(3p_m+2\rho_m+3\rho_m
p_m)\right]
\end{equation}

which is identical to the equation \eqref{mod accln eqn} with
$\Lambda(t)_{\text{eff}}\sim \rho_\Lambda=0$. Hence in this case,
the field equation and the corresponding energy conservation
equation become

\begin{equation}
\frac{{\dot{a}}^2}{a^2}=\frac{8\pi G}{3}{{\rho}_m}_{\text{eff}}
\end{equation}

and

\begin{equation}
\dot{\rho_m} + 3\frac{\dot{a}}{a}(\rho_m+p_m)=0
\end{equation}

However, any other form of $S_H$ will not result in the above set
of equations and the cosmological implications will definitely be
something else.\\

\section{\normalsize\bf{GUP Corrected Entropy on the Horizon}}

In recent years, a number of studies in general relativity and
modified gravity theories came to surface due to the discovery of
different aspects of black hole solutions. Black holes are
thermodynamic objects with well defined entropy. Generally, the
Bekenstein Hawking entropy (\cite{Bek1},\cite{Bek2},\cite{Bek3})

\begin{equation}
S_{BH}=\frac{A}{4{l_p}^2}
\end{equation}

is chosen for the same. Here $A$ is the surface area of the sphere
with the Hubble horizon $r_H=\frac{1}{H}$ and
$l_p=\sqrt{\frac{G\hbar}{c^3}}\simeq 10^{-35}$m is the Planck
length. With $A=4\pi{r_H}^2$, we can write

\begin{equation}\label{Bek entropy}
S_{BH}=\frac{\pi{r_H}^2}{{l_p}^2}
\end{equation}

Instead of a flat universe, if we choose a non-flat universe, then
the apparent horizon $r_A=\frac{1}{\sqrt{H^2+\frac{k}{a^2}}}$
should be used as the horizon radius instead of the Hubble
horizon. Corrections in this entropy formula were needed to
accommodate the newly emerging physics from string theory and loop
quantum gravity (LQG). Several of these theories predicted quantum
corrections to the entropy-area relation
(\cite{Kaul}-\cite{Akbar})

\begin{equation}
S_{QG}=\frac{A}{4{l_p}^2}+C_0~{\text{ln}}\left(\frac{A}{4{l_p}^2}\right)+{\sum_{n=1}}^{\infty}
C_n\left(\frac{A}{4{l_p}^2}\right)^{-n}
\end{equation}

where the coefficients $C_n$ are model dependent parameters.
Recent rigorous calculations from LQG has fixed the value of
$C_0=-1/2$ \cite{Meissner}. On the other hand, Mead \cite{Mead}
first pointed out that Heisenberg uncertainty principle could be
affected by gravity. Later, a considerable amount of effort had
been put to the modified commutation relations between position
and momenta commonly known as the Generalized Uncertainty
Principle (GUP) from different perspectives of quantum aspects of
gravity. All these studies eventually led to the GUP corrected
entropy form (\cite{Barun1}-\cite{Barun2})

\begin{equation}
S_{GUP}=\frac{A}{4{l_p}^2}+\frac{\sqrt{\pi}\alpha_0}{4}\sqrt{\frac{A}{4{l_p}^2}}
-\frac{\pi{\alpha_0}^2}{64}~{\text{ln}}\left(\frac{A}{4{l_p}^2}\right)+O({l_p}^3)
\end{equation}

Here $\alpha_0$ is a dimensionless constant prescribed in the
deformed commutation relations \cite{Das2}. The leading
contribution of this new entropy function lies in its second term
$\sim\sqrt{Area}$. This is an extra term to the already existing
logarithmic correction to entropy derived from the quantum gravity
effects. Due to the difference in the leading order correction
term, the underlying nature of such model needs to be investigated
in four dimensional Einstein's gravity as well as in higher
dimensional modified theories of gravity. Based on many
similarities between the black hole horizon and cosmological
horizon and on the assumption that the universe should be
described by the quantum language, we employ this newly obtained
GUP corrected entropy of the black hole horizon as the entopy of
the cosmological horizon in the natural unit system

\begin{equation}
S_Q=\frac{A}{4}+\frac{\sqrt{\pi}\alpha_0}{4}\sqrt{\frac{A}{4}}
-\frac{\pi{\alpha_0}^2}{64}~{\text{ln}}\left(\frac{A}{4}\right)
\end{equation}

which on further calculation becomes

\begin{equation}
S_Q=S_{BH}\left[1+\frac{\alpha_0
H}{4}-\frac{{\alpha_0}^2H^2}{64}~{\text{ln}}\left(\frac{\pi}{H^2}\right)\right]
\end{equation}

Here $S_{BH}=\frac{\pi}{H^2}$. The novelty of this expression is
that, when $\alpha_0=0$, then $S_Q$ becomes $S_{BH}$.\\

\subsection{\normalsize\bf{Consequences of GUP Corrected
Entropy into the Holographic Equipartition Law}}

Here, we apply the GUP corrected entropy function $S_Q$ into the
Holographic Equipartition Law, i.e, we consider that

\begin{equation}\label{Quantum entropy}
S_H=S_Q=S_{BH}\left[1+\frac{\alpha_0
H}{4}-\frac{{\alpha_0}^2H^2}{64}~{\text{ln}}\left(\frac{\pi}{H^2}\right)\right]
\end{equation}

Substituting this new form of $S_H$ in \eqref{new accln eqn}, we
have

\begin{equation}\label{accl eq from HEL}
\frac{\ddot{a}}{a}=-\frac{4\pi}{3}\left[(\rho_m+3p_m)+\frac{1}{\lambda}(3p_m+2\rho_m+3\rho_m
p_m)\right]+\left[\frac{{\alpha_0}^2H^4}{64}~{\text{ln}}\left(\frac{\pi}{H^2}\right)-\frac{{\alpha_0}H^3}{4}\right]
\end{equation}

The extra driving term appearing in the right side of the equation
needs to be positive for the current cosmic acceleration.\\

In the brane world gravity, the field equations together with the
continuity equation then become

\begin{equation}
\frac{{\dot{a}}^2}{a^2}=\frac{8\pi}{3}{{\rho}_m}_{\text{eff}}+f_\alpha(H)
\end{equation}

\begin{equation}\label{final accl eq}
\frac{\ddot{a}}{a}=-\frac{4\pi}{3}\left[(\rho_m+3p_m)+\frac{1}{\lambda}(2\rho_m+3p_m+3\rho_m
p_m)\right]+f_\alpha(H)
\end{equation}

and

\begin{equation}\label{new cont eq}
\dot{\rho_m}+3\frac{\dot{a}}{a}(\rho_m+p_m)=-\frac{3\dot{f_\alpha(H)}}{8\pi}
\end{equation}

where the extra term $f_\alpha(H)$ is given by

\begin{equation}\label{extra driving term}
f_\alpha(H)=\frac{{\alpha_0}^2H^4}{64}~{\text{ln}}\left(\frac{\pi}{H^2}\right)-\frac{{\alpha_0}H^3}{4}
\end{equation}

Let us now discuss about the evolution of this extra driving terms
from the entropy function \eqref{Quantum entropy} and the
acceleration equation \eqref{final accl eq}. Equation \eqref{final
accl eq} is the final equation incorporating all three
corrections. As $S_{BH}$ is positive, hence the following
restriction is to be obeyed by the parameters for $S_Q$ to be
positive

\begin{equation}\label{SQ positive}
\left[\frac{\alpha_0H}{16}~{\text{ln}}\left(\frac{\pi}{H^2}\right)-1\right]<\frac{4}{\alpha_0H}
\end{equation}

Again for the current cosmic acceleration

\begin{equation}\label{cosmic accl}
0<f_\alpha(H)=\frac{{\alpha_0}H^3}{4}\left[\frac{\alpha_0H}{16}~{\text{ln}}\left(\frac{\pi}{H^2}\right)
-1\right]
\end{equation}

Hence it is clear from \eqref{SQ positive} and \eqref{cosmic accl}
that

\begin{equation}\label{cosmo term1}
f_\alpha(H)<H^2
\end{equation}

A similar constraint can be derived from the study of the
Generalized Second Law of Thermodynamics (GSLT) as presented in
the following subsection.\\

\subsection{\normalsize\bf{Generalized Second Law of Thermodynamics (GSLT)}}

Here we shall discuss the GSLT in the current prescription.
Considering $S_T$ as the total entropy of the universe, one can
write

\begin{equation}
\dot{S_T}=\dot{S_Q}+\dot{S_I}
\end{equation}

where $S_I$ is the entropy of matter inside the horizon. From
\eqref{Quantum entropy}, we can write

\begin{equation}\label{quant entropy change}
\dot{S_Q}=\dot{S_{BH}}\left[1-\left(\frac{{\alpha_0}^2H^2}{64}-\frac{{\alpha_0}H}{8}\right)\right]
\end{equation}

where

\begin{equation}
\dot{S_{BH}}=\frac{d}{dt}\left(\frac{\pi}{H^2}\right)=-\frac{2\pi\dot{H}}{H^3}
\end{equation}

Since $\dot{S_{BH}}>0$, to satisfy $\dot{S_Q}>0$, the following
restriction needs to be obeyed

\begin{equation}\label{restriction1}
\left(\frac{{\alpha_0}^2H^2}{64}-\frac{{\alpha_0}H}{8}\right)<1
\end{equation}

In order to obtain the rate of change of entropy of the matter
inside the horizon, we consider the Gibbs' equation
(\cite{Abdalla},\cite{Pavon})

\begin{equation}
T_IdS_I=dE_I+p_TdV
\end{equation}

where $V$ is the volume inside the horizon and $E_I=\rho_TdV$
stands for the internal energy. The temperature of the matter
$T_I$ inside the horizon has been assumed to be equivalent to the
horizon temperature $T=\frac{H}{2\pi}$. In absence of any dark
energy component, this equation takes the form

\begin{eqnarray*}
T\dot{S_I}=\left[\dot{\rho_m}+3\frac{\dot{a}}{a}(\rho_m+p_m)\right]V
\end{eqnarray*}
\begin{equation}
=-\frac{3\dot{f_\alpha(H)V}}{8\pi}
\end{equation}

where we have used the modified continuity equation \eqref{new
cont eq} to obtain the expression of $\dot{S_I}$. Taking time
derivative of \eqref{extra driving term} and using the expression
of horizon temperature $T$, one can yield

\begin{equation}\label{Int entropy change}
\dot{S_I}=\dot{S_{BH}}\left[\frac{{\alpha_0}^2H^2}{32}~{\text{ln}}\left(\frac{\pi}{H^2}\right)
-\frac{3{\alpha_0}H}{8}-\frac{{\alpha_0}^2H^2}{64}\right]
\end{equation}

Thus from \eqref{quant entropy change} and \eqref{Int entropy
change}, the rate of change of total entropy of the universe
becomes

\begin{equation}
\dot{S_T}=\dot{S_{BH}}\left[1-\left\{\frac{\alpha_0H}{4}+\frac{{\alpha_0}^2H^2}{32}
-\frac{{\alpha_0}^2H^2}{32}~{\text{ln}}\left(\frac{\pi}{H^2}\right)\right\}\right]
\end{equation}

Again as $\dot{S_{BH}}>0$, to satisfy $\dot{S_T}>0$, the following
condition must be attained

\begin{equation}\label{restriction2}
\left[\frac{\alpha_0H}{4}+\frac{{\alpha_0}^2H^2}{32}
-\frac{{\alpha_0}^2H^2}{32}~{\text{ln}}\left(\frac{\pi}{H^2}\right)\right]<1
\end{equation}

From \eqref{restriction1} and \eqref{restriction2}, one can easily
derive

\begin{equation}\label{cosmo term2}
f_{\alpha}(H)>\frac{H^2}{2}
\end{equation}

Thus, we attain a very interesting result from \eqref{cosmo term1}
and \eqref{cosmo term2}

\begin{equation}
\frac{H^2}{2}<f_{\alpha}(H)<H^2
\end{equation}

Following the arguments of \cite{Komatsu4} as for the
observational constraint $\dot{H}<0$ \cite{Krishna}, one can
assume $H_0$ to be the minimum value for $H$ and arrive at a
stricter constraint

\begin{eqnarray*}
\frac{{H_0}^2}{2}<f_\alpha(H)<{H_0}^2
\end{eqnarray*}
\begin{equation}
\Rightarrow O(f_\alpha(H))~\lesssim~O({H_0}^2)
\end{equation}

This result is analogous to the one presented in both
\cite{Komatsu3} and \cite{Komatsu4}, though in the former study, a
mathematical condition was imposed to obtain similar restriction
while in the later, it evolved through the validity of the GSLT.
Further probing into the standard $\Lambda$CDM model, we obtain
$\Lambda=3{H_0}^2\Omega_\Lambda$. This implies that

\begin{equation}
O\left(\frac{\Lambda}{3}\right)=O({H_0}^2\Omega_\Lambda)
\end{equation}

As from Planck (2015) results \cite{Ade}, $\Omega_\Lambda=0.692$,
which is of order one. This yields to

\begin{equation}
O\left(\frac{\Lambda}{3}\right)\simeq O({H_0}^2)
\end{equation}

Thus the order of the extra driving term in the acceleration
equation becomes equivalent to the order of the cosmological
constant term. This result however seems to be model-independent
as the positive brane tension did not play any significant role in
deriving the analogy.\\

\section{\normalsize\bf{Discussions}}

In the present work, our aim was to study the cosmic evolution in
the Brane world gravity with the help of the Holographic
Equipartition Law. We have applied quantum corrected form of the
entropy function derived from the Generalized Uncertainty
Principle in the Holographic Equipartition Law to derive the
modified cosmological equations in a homogeneous, isotropic and
spatially flat 3-brane embedded in a five dimensional bulk. The
novelty of the study lies in $\sqrt{Area}$ form of the entropy
function. It was noticed that the acceleration equation contains
an extra driving term of order consistent with the order of the
cosmological constant. A similar constraint was obtained assuming
the validity of GSLT. The study remained to be model independent
and the positive brane tension did not play any crucial role for
the attained result. However, it should be understood that our aim
was not to verify the GSLT in the modified gravity theory. Rather
we were interested in the evolution of the extra driving term
appearing in the acceleration equation due to imposition of the
holographic equipartition law for a specific GUP corrected entropy
function whose leading order term is different from the existing
forms. This may reflect new light to the studies of the
cosmological constant
problem in modified gravity theories.\\\\

{\bf Acknowledgement}:\\

The author is thankful to IUCAA, Pune, for their warm hospitality
and excellent research facilities where part of the work has been
done during a visit under the Associateship Programme.\\\\

\end{document}